\documentstyle[12pt]{article}

\begin{document}

\title{Gravity as Lorentz Force}

\author{Yury M. Zinoviev\thanks{This work is supported in part by the Russian
Foundation for Basic Research (Grant No. 00 -- 01 -- 00083)}\\ Steklov
Mathematical Institute,\\ Gubkin St. 8, Moscow 117966, GSP -- 1, Russia \\
e -- mail: zinoviev@mi.ras.ru }

\date{}

\maketitle

\vskip 1cm

\noindent {\bf Abstract}. The main object of the proposed theory is not a
pseudometric, but a symmetric affine connection on the Minkowski space. The 
coefficients of this connection have one upper and two lower indices. 
These coefficiens are symmetric with respect to the permutation of the 
lower indices. We identify the convolution of the connection coefficients 
with the vector -- potential of the electromagnetic field. Then the gravity 
is the Lorentz force of this electromagnetic field.

\vskip 1cm

\section{Introduction}

\noindent The idea of the relativistic gravitation theory was proposed by
Poincar\'e \cite{1} :

"In the paper cited Lorentz found it necessary to supplement his hypothesis
so that the relativity postulate could be valid for other forces besides the
electromagnetic ones. According to his idea, owing to Lorentz transformation
(and therefore owing to the translational movement) all forces behave like
electromagnetic.

It turned out to be necessary to consider this hypothesis more attentively
and to study the changes it makes in the gravity laws in particular. First
of all, it enables us to suppose that the gravity forces propagate not 
instantly, but at the light velocity. One could think that it is enough to
reject such a hypothesis, for Laplace has shown that it can not take place.
But in fact the effect of this propagation is largely balanced by some other
circumstance, so there is no any contradiction between the law proposed and 
the astronomical observations.

Is it possible to find a law satisfying the condition stated by Lorentz and
at the same time coming to the Newton law in all the cases when the velocities
of the celestial bodies are small enough to neglect their squares (and also
the products of the accelerations and the distance) with respect to the 
square of the velocity of light?"

Poincar\'e found that the mathematical solution of the problem is not unique.
It is easy to solve the Poincar\'e problem by making use of the physical
reasons. The form of the Newton gravity law coincides with the form of the
Coulomb law describing the interaction between two oppositely charged bodies. 
The relativistic form of the Coulomb law is well -- known. It is the
Lorentz force. Thus the relativistic form of the gravity law must be the same.

It is possible to solve the derived equations by using a computer.  We
consider a simple problem of the interaction of two bodies when the mass
of one body is equal to zero. It is a problem of the light propagation in
the gravity field of one body. The received exact solution of this problem
describes all effects predicted by the general relativity: the distortion
of the light beams in the gravity field, the light motion along the closed
trajectory in the gravity field, etc. If the Mercury mass is considered small,
it is possible to calculate the shift of Mercury perihelion for a hundred
years. It turns out to be 45". The general relativity predicts 43".

In this paper we do not consider the radiation effects. It seems that these
effects lead to the non -- stability of the orbits of the celestial bodies.
These effects need computer calculations. It seems that the study of these
effects requires not electrodynamics, but quantum electrodynamics.

In the general relativity the equation of motion in the gravity field is
the geodesic equation where the parameter is the proper time corresponding
to the pseudometric defining the gravity field. It is possible to obtain the
Newton gravity law from this equation only when one of two bodies is at rest.
For an arbitrary case it is needed the geodesic equation with the parameter
not connected to the pseudometric and with the connection coefficients
which are not the Christoffel's symbols (the connection is not Riemannian and
it is not compatible with any pseudometric). Therefore it seems natural to
consider the generalization of general relativity when not a pseudometric
but a symmetric affine connection on the Minkowski space is given. As the 
equations of motion we consider all possible equations of geodesic type. 
The parameter of these equations is the proper time corresponding to the 
Minkowski pseudometric. It turns out that the unique Lagrangian equation 
among these equations is the equation of motion under the action of 
Lorentz force. We come back to Poincar\'e idea.

Let us consider the equation of field. In the book (\cite{2} , Chapter 7)
all theories are examined where gravitation potential is taken in the form
of scalar, vector and at last symmetric tensor fields. All such theories
have their defects. We consider the gravitation potential as the coefficient
of the symmetric affine connection on the Minkowski space. This coefficient 
has three indices and it is not a tensor. It is possible to consider this 
coefficient as Yang -- Mills field. The curvature tensor corresponding to 
this connection is the Yang -- Mills field strength tensor. Following 
Einstein, we study the convolutions of the curvature tensor. The
curvature tensor has one upper and three lower indices. It is antisymmetric
in respect of the  permutation of two lower indices. Therefore two 
convolutions of the curvature tensor are possible. One of these convolutions
is the trace of the Yang -- Mills field strength tensor. It coincides with the
electromagnetic field strength tensor. The vector -- potential of this
electromagnetic field is the convolution of the symmetric affine connection
coefficients. The symmetric affine connection coefficient has one upper
and two lower indices. It is symmetric with respect to the permutation of
the lower indices. Hence the convolution of the coefficients of the
symmetric affine connection is unique. The second convolution of the
curvature tensor is the Ricci tensor. The antisymmetric part of Ricci tensor
is proportional to above electromagnetic field strength tensor. If the
symmetric affine connection is Riemannian (compatible with some 
pseudometric), then the electromagnetic field strength tensor is equal to
zero and the Ricci tensor is symmetric with respect to the permutation of its
indices. The symmetric affine connection coefficients satisfy the wave
equations. The substitution of the first convolution of the curvature
tensor into the equation of motion under the action of the Lorentz force
yields the relativistic Newton gravity law.

In the second section we study the relativistic Newton gravity law. The
third section is devoted to the relation between the geodesic equation and
the Newton gravity law. In the fourth section the cuvature tensor of the
symmetric affine connection on the Minkowski space is considered.

\section{Relativistic Newton Gravity Law}
\setcounter{equation}{0}

\noindent Let two bodies have the masses $m_{k}$, $k = 1,2$, and move
along the world lines $x_{k}^{\mu }(t)$, $k = 1,2$, $\mu = 0,...,3$,
$x_{k}^{0}(t) = ct$. The gravitation interaction of these bodies is
given by the Newton gravity law
\begin{equation}
\label{2.1}
\frac{d^{2}x_{k}^{i}}{dt^{2}} = - 
\frac{\partial U(x_{k};\hat{x}_{k})}{\partial x_{k}^{i}}, 
\end{equation}  
\begin{equation}
\label{2.2}
\sum_{j = 1}^{3} \left( \frac{\partial}{\partial x^{j}} \right)^{2} U(x;y) =
4\pi m_{y}G\delta ({\bf x} - {\bf y}(t))
\end{equation}
where $i = 1,2,3$; $k = 1,2$; $U(x_{1};\hat{x}_{1}) = U(x_{1};x_{2})$,
$U(x_{2};\hat{x}_{2}) = U(x_{2};x_{1})$ and the gravitation constant
\begin{equation}
\label{2.3}
G = (6,673 \pm 0,003)\cdot 10^{- 11} \, \, \, m^{3} \cdot kg^{- 1} \cdot
s^{- 2}.
\end{equation}

The form of the law (\ref{2.1}), (\ref{2.2}) coincides with the form of 
the Coulomb law for the electric interaction of two oppositely charged bodies. 
The relativistic form of the Coulomb law is the movement under the action of
the Lorentz force. Therefore the relativistic form of the Newton gravity law 
is the movement under the action of the Lorentz force
\begin{equation}
\label{2.4}
c \frac{d^{2}x_{k}^{\mu }}{ds_{k}^{2}} =  - \frac{1}{c} \eta^{\mu \mu }
\sum_{\nu = 0}^{3} F_{\mu \nu }(x_{k};\hat{x}_{k} )
\frac{dx_{k}^{\nu }}{ds_{k}},
\end{equation}
\begin{equation}
\label{2.5}
F_{\mu \nu }(x;y) = \frac{\partial A_{\nu }(x;y)}{\partial x^{\mu }} -
\frac{\partial A_{\mu }(x;y)}{\partial x^{\nu }},
\end{equation}
\begin{equation}
\label{2.6}
\sum_{\nu = 0}^{3} \eta^{\nu \nu } 
\left( \frac{\partial }{\partial x^{\nu }} \right)^{2} A_{\mu }(x;y) =
\eta_{\mu \mu } 4\pi m_{y} G \frac{dy^{\mu }(\frac{x^{0}}{c})}{dx^{0}} 
\delta ({\bf x} - {\bf y}(\frac{x^{0}}{c})),
\end{equation} 
\begin{equation}
\label{2.7}
\frac{dx^{\mu }}{ds} = \frac{dt}{ds} \frac{dx^{\mu }}{dt}, \, \, \, \,
\frac{d^{2}x^{\mu }}{ds^{2}} = \frac{dt}{ds} \frac{d}{dt}
\left( \frac{dt}{ds} \frac{dx^{\mu }}{dt} \right),
\end{equation}
\begin{equation}
\label{2.8}
\frac{dt}{ds} = 
c^{- 1}(1 - \frac{1}{c^{2}} |\frac{d{\bf x}}{dt} |^{2})^{- 1/2}
\end{equation}
where $\mu = 0,...,3$; $k = 1,2$; 
$|{\bf v}|^{2} = \sum_{i = 1}^{3} (v^{i})^{2}$; the $4\times 4$ matrix
$\eta_{\mu \nu} $ is diagonal and 
$\eta_{00} = - \eta_{11} = - \eta_{22} = - \eta_{33} = 1$; the
$4\times 4$ matrix $\eta^{\mu \nu}$ is the inverse matrix of $\eta_{\mu \nu} $.

Let the second body be at rest: $x_{2}^{i}(t) = 0$, $i = 1,2,3$. Let
$A_{i}(x) = 0$, $i = 1,2,3$, and $A_{0}({\bf x})$ be the solution of the
equation (\ref{2.6}) for $\mu = 0$, $y^{i}(t) = 0$, $i = 1,2,3$. 
$A_{0}({\bf x})$ does not depend on the variable $x^{0}$. Then the 
equations (\ref{2.4}) for $k = 1$ have the following form
\begin{equation}
\label{2.9}
c\frac{d}{dt} 
\left( (1 - \frac{1}{c^{2}} |\frac{d{\bf x}_{1}}{dt} |^{2})^{- 1/2} \right)
= \frac{1}{c} \sum_{i = 1}^{3} 
\frac{\partial A_{0}({\bf x}_{1})}{\partial x_{1}^{i}}
\frac{dx_{1}^{i}}{dt}, 
\end{equation}
\begin{equation}
\label{2.10}
\frac{d}{dt} 
\left( (1 - \frac{1}{c^{2}} |\frac{d{\bf x}_{1}}{dt}|^{2})^{- 1/2} 
\frac{dx_{1}^{i}}{dt} \right) = 
\frac{\partial A_{0}({\bf x}_{1})}{\partial x_{1}^{i}}
\end{equation}
where $i = 1,2,3$. The following identity is valid
\begin{equation}
\label{2.11}
c\frac{d}{dt} 
\left( (1 - \frac{1}{c^{2}} |\frac{d{\bf x}_{1}}{dt}|^{2})^{- 1/2} \right)
= \frac{1}{c} \sum_{i = 1}^{3} \frac{dx_{1}^{i}}{dt}
\frac{d}{dt} 
\left( (1 - \frac{1}{c^{2}} |\frac{d{\bf x}_{1}}{dt}|^{2})^{- 1/2} 
\frac{dx_{1}^{i}}{dt} \right).
\end{equation}
This identity implies that the equation (\ref{2.9}) is the consequence of 
the equations (\ref{2.10}). The comparison of the equations (\ref{2.2})
and (\ref{2.6}) for $y^{i}(t) = 0$ gives $A_{0}({\bf x}) = - U({\bf x})$.
Hence if $|\frac{dx_{1}}{dt} |^{2}$ is small with respect to $c^{2}$,
the equations (\ref{2.10}) coincide with the equations (\ref{2.1}) for
$k = 1$.

Let us consider the general case when two bodies move. By using the identity
(\ref{2.11}) it is easy to prove that the equations (\ref{2.4}) for
$\mu = 0$ are the consequence of the equations (\ref{2.4}) for
$\mu = 1,2,3$. Since $x_{k}^{0}(t) = ct$ the equations (\ref{2.4}) -- 
(\ref{2.6}) may be written in following form
\begin{eqnarray}
\label{2.12}
\frac{d}{dt} 
\left( (1 - \frac{1}{c^{2}} |\frac{d{\bf x}_{k}}{dt}|^{2})^{- 1/2} 
\frac{dx_{k}^{i}}{dt} \right) =
\frac{\partial A_{0}(x_{k}(t);\hat{x}_{k} )}{\partial x_{k}^{i}} -
\frac{1}{c} \frac{\partial A_{i}(x_{k}(t);\hat{x}_{k} )}{\partial t} +
\nonumber \\
\frac{1}{c} \sum_{j = 1}^{3} \left( 
\frac{\partial A_{j}(x_{k}(t);\hat{x}_{k} )}{\partial x_{k}^{i}} -
\frac{\partial A_{i}(x_{k}(t);\hat{x}_{k} )}{\partial x_{k}^{j}} \right)
\frac{dx_{k}^{j}}{dt},
\end{eqnarray}
\begin{equation}
\label{2.13}
\left( \frac{1}{c^{2}} \left( \frac{\partial}{\partial t} \right)^{2} -
\sum_{j = 1}^{3} \left( \frac{\partial }{\partial x^{j}} \right)^{2} \right)
A_{0}(x;y) = 4\pi m_{y}G \delta ({\bf x} - {\bf y}(t)),
\end{equation}
\begin{equation}
\label{2.14}
\left( \frac{1}{c^{2}} \left( \frac{\partial}{\partial t} \right)^{2} -
\sum_{j = 1}^{3} \left( \frac{\partial }{\partial x^{j}} \right)^{2} \right)
A_{i}(x;y) = - \frac{4\pi}{c} m_{y}G \frac{dy^{i}}{dt}
\delta ({\bf x} - {\bf y}(t)).
\end{equation}
If we formally allow $c$ to tend to the infinity, the equations (\ref{2.12})
-- (\ref{2.14}) coincide with the equations (\ref{2.1}), (\ref{2.2}). In
order to study the limit $c \rightarrow \infty $ we need to consider the
behaviour of the solutions of the equations (\ref{2.12}) -- (\ref{2.14})
when $c \rightarrow \infty $. It will be done below for the particular
case.

Let us verify that the equations (\ref{2.4}) -- (\ref{2.6}) are Lorentz
covariant. For the world line $y^{\mu }(t)$, $y^{0}(t) = ct$, we define
the current
\begin{equation}
\label{2.15}
j^{\mu }(x;y) = c\frac{dy^{\mu }(\frac{x^{0}}{c})}{dx^{0}}
\delta ({\bf x} - {\bf y}(\frac{x^{0}}{c})).
\end{equation}
The world line $y^{\mu }(t)$, $y^{0}(t) = ct$, is called timelike if the
vector $\frac{dy^{\mu }(t)}{dt}$ lies in the upper light cone
\begin{equation}
\label{2.16}
|\frac{d{\bf y}(t)}{dt} | < c.
\end{equation}
{\bf Lemma 2.1.} {\it If the world line} $y^{\mu }(t)$, $y^{0}(t) = ct$,
{\it is timelike, then the current} (\ref{2.15}) {\it satisfies the
covariance relation}
\begin{equation}
\label{2.17}
j^{\mu }(\Lambda x;\Lambda y) = \sum_{\nu = 0}^{3} \Lambda_{\nu }^{\mu }
j^{\nu }(x;y)
\end{equation}
{\it for an arbitrary matrix} $\Lambda_{\nu }^{\mu }$ {\it from Lorentz
group.}

\noindent {\it Proof.} The definition (\ref{2.15}) may be rewritten as
\begin{equation}
\label{2.18}
j^{\mu }(x;y) = \int \delta (x - y(t)) \frac{dy^{\mu }(t)}{dt} dt,
\end{equation}
\begin{equation}
\label{2.19}
\delta (x - y(t)) = \delta (x^{0} - y^{0}(t)) \delta ({\bf x} - {\bf y}(t)).
\end{equation}
Let $4\times 4$ matrix $\Lambda_{\nu }^{\mu }$ belong to the Lorentz group.
By making use of the relation (\ref{2.18}) we have
\begin{equation}
\label{2.20}
\sum_{\nu = 0}^{3} \Lambda_{\nu }^{\mu } j^{\nu }(\Lambda^{- 1} x;y) =
\int \delta (x - \Lambda y(t)) \frac{d}{dt} (\Lambda y(t))^{\mu } dt,
\end{equation}
\begin{equation}
\label{2.21}
(\Lambda y(t))^{\mu } = \sum_{\nu = 0}^{3} \Lambda_{\nu }^{\mu } y^{\nu }(t).
\end{equation}
Since the vector $\frac{dy^{\mu }(t)}{dt}$ lies in the upper light cone,
the vector $(\Lambda \frac{dy(t)}{dt})^{\mu }$ lies in the upper light
cone too.

Let the relation
\begin{equation}
\label{2.22}
ct_{1} = \sum_{\nu = 0}^{3} \Lambda_{\nu }^{0} y^{\nu }(t(t_{1}))
\end{equation}
define the mapping $t(t_{1})$. The mapping $t(t_{1})$ has positive
derivative
\begin{equation}
\label{2.23}
\frac{dt}{dt_{1}} = c ((\Lambda \frac{dy}{dt} (t(t_{1})))^{0})^{- 1}
\end{equation}
as the vector $(\Lambda \frac{dy(t)}{dt} )^{\mu }$ lies in the upper light
cone.

Taking the integration variable $t_{1}$ in the right hand side of the 
relation (\ref{2.20}) we get
\begin{equation}
\label{2.24}
\sum_{\nu = 0}^{3} \Lambda_{\nu }^{\mu } j^{\nu }(\Lambda^{- 1} x;y) =
\int \delta (x - \Lambda y(t(t_{1}))) \frac{d}{dt_{1}}
(\Lambda y(t(t_{1})))^{\mu } dt_{1}.
\end{equation}
The world line $(\Lambda y(t(t_{1})))^{\mu }$ has the property (\ref{2.22}).
In virtue of the world line $y^{\mu }(t)$ is timelike, the world line
$(\Lambda y(t(t_{1})))^{\mu }$ is timelike too. The relations (\ref{2.18}),
(\ref{2.24}) imply
\begin{equation}
\label{2.25}
\sum_{\nu = 0}^{3} \Lambda_{\nu }^{\mu } j^{\nu }(\Lambda^{- 1} x;y) =
j^{\mu }(x;\Lambda y).
\end{equation}
The substitution $\Lambda x$ for $x$ in the relation (\ref{2.25}) gives
the relation (\ref{2.17}). The lemma is proved.

Lemma 2.1 implies the Lorentz covariance of the equations (\ref{2.4}) -- 
(\ref{2.6}). Any solution of the equation (\ref{2.6}) has the following
form
\begin{equation}
\label{2.26}
A_{\mu }(x;y) = \eta_{\mu \mu} 4\pi m_{y}G \int E(x - x_{1})
\frac{dy^{\mu }(\frac{x_{1}^{0}}{c})}{dx_{1}^{0}}
\delta ({\bf x}_{1} - {\bf y}(\frac{x_{1}^{0}}{c} )) d^{4}x_{1}
\end{equation}
where the distribution $E(x) \in S^{\prime }({\bf R}^{4})$ satisfies
the equation
\begin{equation}
\label{2.27}
\sum_{\nu = 0}^{3} \eta^{\nu \nu } \left( \frac{\partial }{\partial x^{\nu }} 
\right)^{2} E(x) = \delta (x).
\end{equation}
The solution (\ref{2.26}) is Lorentz covariant if the distribution $E(x)$
is Lorentz invariant.

\noindent {\bf Lemma 2.2.} {\it Any Lorentz invariant solution of the
equation} (\ref{2.27}) {\it has the following form}
\begin{eqnarray}
\label{2.28}
E(x) = (2\pi )^{- 1} \theta (x^{0}) \delta ((x^{0})^{2} - |{\bf x}|^{2}) +
c_{1}\epsilon (x^{0}) \delta ((x^{0})^{2} - |{\bf x}|^{2}) + \nonumber \\
c_{2}\left( ((x^{0} + i0)^{2} - |{\bf x}|^{2})^{- 1} +
((x^{0} - i0)^{2} - |{\bf x}|^{2})^{- 1} \right) + c_{3}
\end{eqnarray}
{\it where} $c_{1}$, $c_{2}$, $c_{3}$ {\it are the arbitrary constants; the
step function}
\begin{equation}
\label{2.29}
\theta (x^{0}) = \left\{ 1, \hskip 1cm x^{0} > 0 \atop
{0, \hskip 1cm x^{0} < 0} \right. ;
\end{equation}
{\it the sign function} $\epsilon (x^{0}) = \theta (x^{0}) - \theta (x^{0})$
{\it and} $\left( (x^{0} \pm i0)^{2} - |{\bf x}|^{2} \right)^{- 1} $
{\it are the disrtibution boundary values of the holomorphic functions}
$\left( (x^{0} \pm i\epsilon )^{2} - |{\bf x}|^{2} \right)^{- 1} $.

\noindent {\it Proof.} Due to (\cite{3}, Chapter 5, Section 30) the 
distribution (\ref{2.28}) satifies the equation (\ref{2.27}). It is Lorentz
invariant. Let us prove that the equation (\ref{2.27}) has no other
Lorentz invariant solutions. It is sufficient to prove that any Lorentz
invariant distribution satisfying the wave equation
\begin{equation}
\label{2.30}
\sum_{\nu = 0}^{3} \eta^{\nu \nu } \left( \frac{\partial }{\partial x^{\nu }} 
\right)^{2} f(x) = 0
\end{equation}
has the following form
\begin{equation}
\label{2.31}
f(x) = 
c_{1}\epsilon (x^{0}) \delta ((x^{0})^{2} - |{\bf x}|^{2}) + 
c_{2}\left( ((x^{0} + i0)^{2} - |{\bf x}|^{2})^{- 1} +
((x^{0} - i0)^{2} - |{\bf x}|^{2})^{- 1} \right) + c_{3}
\end{equation}
where $c_{i}$, $i = 1,2,3$ are the arbitrary constants.

The first distribution in the right hand side of the equality (\ref{2.31})
is odd with respect to the reflection: $f(- x) = - f(x)$. The two last 
distributions in the right hand side of the equality (\ref{2.31}) are even
with respect to the reflection: $f(- x) = f(x)$.

Let the distribution $f(x)$ be Lorentz invariant and odd with respect
to the reflection. Its Fourier transform $F[f](x)$ has the same properties and
satisfies the equation
\begin{equation}
\label{2.32}
((x^{0})^{2} - |{\bf x}|^{2}) F[f](x) = 0.
\end{equation}
Due to \cite{4} for any Lorentz invariant odd with respect to the reflection
distribution $F[f](x) \in S^{\prime }({\bf R}^{4})$ there is the
distribution $g(t) \in S^{\prime }({\bf R})$ such that
\begin{equation}
\label{2.33}
\int F[f](x) \phi (x) d^{4}x =
\int dt g(t) \int \epsilon (x^{0}) \delta ((x^{0})^{2} - |{\bf x}|^{2} - t)
\phi (x) d^{4}x
\end{equation}
for any test function $\phi (x)$ vanishing with its every derivative at the
point $x = 0$. If the distribution $F[f](x)$ satisfies the equation 
(\ref{2.32}), then the distribution $g(t)$ satisfies the equation
$tg(t) = 0$. Therefore $g(t) = c_{1}^{\prime }\delta (t)$ where 
$c_{1}^{\prime }$ is an arbitrary constant. The substitution of this
solution into the right hand side of the equality (\ref{2.33}) gives
\begin{equation}
\label{2.34}
\int F[f](x) \phi (x) d^{4}x =
c_{1}^{\prime } \int \epsilon (x^{0}) 
\delta ((x^{0})^{2} - |{\bf x}|^{2}) \phi (x) d^{4}x
\end{equation}
for any test function $\phi (x)$ vanishing with its every derivative at the
point $x = 0$. The difference between the distribution in the left hand
side of the equality (\ref{2.34}) and the distribution in the right hand
side of the equality (\ref{2.34}) is Lorentz invariant distribution with
the support at the point $x = 0$
\begin{equation}
\label{2.35}
F[f](x) - c_{1}^{\prime } \epsilon (x^{0}) 
\delta ((x^{0})^{2} - |{\bf x}|^{2}) =
\sum_{k = 0}^{N} a_{k} \left( \sum_{\nu = 0}^{3} \eta^{\nu \nu }
\left( \frac{\partial }{\partial x^{\nu }}\right)^{2} \right)^{k}
\delta (x).
\end{equation}
The left hand side of the equality (\ref{2.35}) is odd with respect to
the reflection and the right hand side of the equality (\ref{2.35}) is even
with respect to the reflection. Hence both sides of the equality (\ref{2.35})
are equal to zero. Then due to (\cite{3}, Chapter 5, Section 30) we have
\begin{equation}
\label{2.36}
f(x) = i(2\pi )^{2} c_{1}^{\prime } \epsilon (x^{0}) 
\delta ((x^{0})^{2} - |{\bf x}|^{2}).
\end{equation}
Thus any Lorentz invariant odd with respect to the reflection satisfying the
wave equation (\ref{2.30}) distribution has the form of the first 
distribution in the right hand side of the equality (\ref{2.31}).

Let the distribution $f(x)$ be Lorentz invariant even with respect to the
reflection and satisfy the wave equation (\ref{2.30}). Then its Fourier
transform $F[f](x)$ is Lorentz invariant even with respect to the reflection
and satifies the equation (\ref{2.32}). Due to \cite{4} there is the
distribution $g(t) \in S^{\prime }({\bf R})$, such that
\begin{equation}
\label{2.37}
\int F[f](x) \phi (x) d^{4}x =
\int dt g(t) \int \delta ((x^{0})^{2} - |{\bf x}|^{2} - t)
\phi (x) d^{4}x
\end{equation}
for any test function $\phi (x)$ vanishing with its every derivative at the
point $x = 0$. If the distribution $F[f](x)$ satisfies the equation 
(\ref{2.32}), then the distribution $g(t)$ satisfies the equation
$tg(t) = 0$. Therefore $g(t) = c_{2}^{\prime }\delta (t)$ where 
$c_{2}^{\prime }$ is an arbitrary constant. The substitution of this
solution into the right hand side of the equality (\ref{2.37}) gives
\begin{equation}
\label{2.38}
\int F[f](x) \phi (x) d^{4}x =
c_{2}^{\prime } \int 
\delta ((x^{0})^{2} - |{\bf x}|^{2}) \phi (x) d^{4}x
\end{equation}
for any test function $\phi (x)$ vanishing with its every derivative at the
point $x = 0$. The difference between the distribution in the left hand
side of the equality (\ref{2.38}) and the distribution in the right hand
side of the equality (\ref{2.38}) is Lorentz invariant distribution with
the support at the point $x = 0$
\begin{equation}
\label{2.39}
F[f](x) - c_{2}^{\prime } \delta ((x^{0})^{2} - |{\bf x}|^{2}) =
\sum_{k = 0}^{N} a_{k} \left( \sum_{\nu = 0}^{3} \eta^{\nu \nu }
\left( \frac{\partial }{\partial x^{\nu }}\right)^{2} \right)^{k}
\delta (x).
\end{equation}
The distributions in the left hand of the equality (\ref{2.39}) satisfy 
the equation (\ref{2.32}). The distribution in the right hand side of the
equality (\ref{2.39}) satisfies the equation (\ref{2.32}) only if
$a_{k} = 0$ for $k > 0$. Then due to (\cite{3}, Chapter 5, Section 30)
we have
\begin{equation}
\label{2.40}
f(x) = 
- 2\pi c_{2}^{\prime }\left( ((x^{0} + i0)^{2} - |{\bf x}|^{2})^{- 1} +
((x^{0} - i0)^{2} - |{\bf x}|^{2})^{- 1} \right) + (2\pi )^{- 4}a_{0}.
\end{equation}
Thus any Lorentz invariant even with respect to the reflection satisfying
the wave equation (\ref{2.30}) distribution has the form of the sum of
two last distributions in the right hand side of the equality (\ref{2.31}).
The lemma is proved.

If we substitute the expression (\ref{2.28}) into the right hand side of
the relation (\ref{2.26}), the electromagnetic field strength will depend
on the whole world line $y^{\mu }(t)$. The equations (\ref{2.4}) are
causal if the electromagnetic field strength depends only on the points
of the world line $y^{\mu }(t)$ lying in the lower light cone with the
origin at the point $x$. Then the support of the distribution $E(x)$
must lie in the upper light cone. It is possible only for
$c_{i} = 0$, $i = 1,2,3$, in the expression (\ref{2.28}).

\noindent {\bf Lemma 2.3.} {\it Any Lorentz covariant solution
of the equations} (\ref{2.5}), (\ref{2.6})
{\it which depends only on the points of the world line} $y^{\mu }(t)$
{\it lying in the lower light cone  with the origin at the point} $x$
{\it has the following form}
\begin{eqnarray}
\label{2.41}
F_{\mu \nu }(x;y) = \eta_{\mu \mu } \eta_{\nu \nu } m_{y}G
\left( c|{\bf x} - {\bf y}(t^{\prime })| -
\sum_{i = 1}^{3} (x^{i} - y^{i}(t^{\prime }))
\frac{dy^{i}(t^{\prime })}{dt^{\prime }} \right)^{- 2} \times \nonumber \\
\left( (x^{\mu } - y^{\mu }(t^{\prime })) 
\frac{d^{2}y^{\nu }(t^{\prime })}{dt^{\prime 2}} -
(x^{\nu } - y^{\nu }(t^{\prime })) 
\frac{d^{2}y^{\mu }(t^{\prime })}{dt^{\prime 2}} \right) + \nonumber \\
\eta_{\mu \mu } \eta_{\nu \nu } m_{y}G
\left( c|{\bf x} - {\bf y}(t^{\prime })| -
\sum_{i = 1}^{3} (x^{i} - y^{i}(t^{\prime }))
\frac{dy^{i}(t^{\prime })}{dt^{\prime }} \right)^{- 3} \times \nonumber \\
\left( \sum_{\alpha = 0}^{3} \eta_{\alpha \alpha }
\left( \frac{dy^{\alpha }(t^{\prime })}{dt^{\prime }} \right)^{2} -
\sum_{\alpha = 0}^{3} \eta_{\alpha \alpha } 
(x^{\alpha } - y^{\alpha }(t^{\prime }))
\frac{d^{2}y^{\alpha }(t^{\prime })}{dt^{\prime 2}} \right) \times \nonumber \\
\left( (x^{\mu } - y^{\mu }(t^{\prime })) 
\frac{dy^{\nu }(t^{\prime })}{dt^{\prime }} -
(x^{\nu } - y^{\nu }(t^{\prime })) 
\frac{dy^{\mu }(t^{\prime })}{dt^{\prime }} \right)
\end{eqnarray}
{\it where the time} $t^{\prime }$ {\it satisfies the equation}
\begin{equation}
\label{2.42}
x^{0} - ct^{\prime } = |{\bf x} - {\bf y}(t^{\prime })|
\end{equation} 
{\it Proof.} The substitution of the expression (\ref{2.28}) with
$c_{i} = 0$, $i = 1,2,3$, into the relation (\ref{2.26}) gives 
\begin{equation}
\label{2.43}
A_{\mu }(x;y) = \eta_{\mu \mu }m_{y}G 
\left( \sum_{\alpha = 0}^{3} \eta_{\alpha \alpha } 
(x^{\alpha } - y^{\alpha }(t^{\prime }))
\frac{dy^{\alpha }(t^{\prime })}{dt^{\prime }} \right)^{- 1}
\frac{dy^{\mu }(t^{\prime })}{dt^{\prime }}
\end{equation}
where the time $t^{\prime }$ satisfies the equation (\ref{2.42}). By 
making use of this equation we have
\begin{equation}
\label{2.44}
\sum_{\alpha = 0}^{3} \eta_{\alpha \alpha } 
(x^{\alpha } - y^{\alpha }(t^{\prime }))
\frac{dy^{\alpha }(t^{\prime })}{dt^{\prime }} =
c|{\bf x} - {\bf y}(t^{\prime })| -
\sum_{i = 1}^{3} (x^{i} - y^{i}(t^{\prime }))
\frac{dy^{i}(t^{\prime })}{dt^{\prime }}.
\end{equation}
The timelike world line $y^{\mu }(t)$ satisfies the inequality (\ref{2.16}).
Therefore the expression (\ref{2.44}) is positive.

The equation (\ref{2.42}) implies
$$
\sum_{\alpha = 0}^{3} \eta_{\alpha \alpha } 
(x^{\alpha } - y^{\alpha }(t^{\prime }))^{2} = 0
$$
\begin{equation}
\label{2.45}
\frac{\partial t^{\prime }}{\partial x^{\nu }} =
\eta_{\nu \nu } (x^{\nu } - y^{\nu }(t^{\prime }))
\left( \sum_{\alpha = 0}^{3} \eta_{\alpha \alpha } 
(x^{\alpha } - y^{\alpha }(t^{\prime }))
\frac{dy^{\alpha }(t^{\prime })}{dt^{\prime }} \right)^{- 1}
\end{equation}
Now by using the definition (\ref{2.5}) and the relations (\ref{2.43}) --
(\ref{2.45}) it is easy to obtain the relation (\ref{2.41}). The lemma
is proved.

The formulae (\ref{2.43}) are called the Li\'enard -- Wiechert potentials
(\cite{5}, Chapter VIII, Section 63). The equations (\ref{2.4}), 
(\ref{2.41}), (\ref{2.42}) are the relativistic Newton gravity law.
If we allow $c$ to tend to the infinity, the equations (\ref{2.4}),
(\ref{2.41}), (\ref{2.42}) coincide with the Newton gravity law (\ref{2.1}),
(\ref{2.2}).

It is easy to generalize the equations (\ref{2.4}) -- (\ref{2.6}) to the
gravitation interaction of many bodies
\begin{equation}
\label{2.46}
c \frac{d^{2}x_{k}^{\mu }}{ds_{k}^{2}} =  - \frac{1}{c} \eta^{\mu \mu }
\sum_{{j = 1}\atop {j \neq k}}^{n} \sum_{\nu = 0}^{3} 
F_{\mu \nu }(x_{k};x_{j}) \frac{dx_{k}^{\nu }}{ds_{k}},
\end{equation}
where $k = 1,...,n$, $\mu = 0,...,3$, the vectors 
$\frac{dx_{k}^{\nu }}{ds_{k}}$, $\frac{d^{2}x_{k}^{\mu }}{ds_{k}^{2}}$
are given by the relations (\ref{2.7}), (\ref{2.8}) and the electromagnetic
field strengthes are given by the relations (\ref{2.41}), (\ref{2.42}).

Let us consider the light propagation in the gravity field. It corresponds
to the equations (\ref{2.4}), (\ref{2.41}), (\ref{2.42}) for $m_{1} = 0$.
Then the second equation (\ref{2.4}) is the following equation
$$
\frac{d}{dt} 
\left( (1 - \frac{1}{c^{2}} |\frac{d{\bf x}_{2}}{dt}|^{2})^{- 1/2} 
\frac{dx_{2}^{\mu }}{dt} \right) = 0.
$$
Hence
$$
(1 - \frac{1}{c^{2}} |\frac{d{\bf x}_{2}}{dt}|^{2})^{- 1/2} 
\frac{dx_{2}^{\mu }}{dt} = const
$$
and
$$
\frac{dx_{2}^{\mu }}{dt} = const.
$$
Therefore the second body moves uniformly and rectilinearly. The equatons
(\ref{2.4}), (\ref{2.41}), (\ref{2.42}) are Lorentz covariant. Let us
choose the coordinate system where the second body is at rest: 
$x_{2}^{0}(t) = ct$, $x_{2}^{i}(t) = 0$, $i = 1,2,3$. Then the equations
(\ref{2.4}) and (\ref{2.41}) imply
\begin{equation}
\label{2.47}
\frac{d}{dt} 
\left( (1 - \frac{1}{c^{2}} |\frac{d{\bf x}_{1}}{dt}|^{2})^{- 1/2} 
\frac{dx_{1}^{i}}{dt} \right) = 
- m_{2}G|{\bf x}_{1}|^{- 3}x_{1}^{i} 
\end{equation}
where $i = 1,2,3$.

Let us find the conservation laws of the equation (\ref{2.47}). Let the
tensor $\epsilon_{ijk}$ be antisymmetric for all indices and 
$\epsilon_{123} = 1$. We define the vector
\begin{equation}
\label{2.48}
M_{k} = \sum_{i,j = 1}^{3} \epsilon_{ijk} 
(x_{1}^{i}\frac{dx_{1}^{j}}{dt} - x_{1}^{j}\frac{dx_{1}^{i}}{dt})
(1 - \frac{1}{c^{2}} |\frac{d{\bf x}_{1}}{dt}|^{2})^{- 1/2}
\end{equation}
where $k = 1,2,3$. It follows from the equations (\ref{2.47}) that
$$
\frac{dM_{k}}{dt} = 0.
$$
The identity (\ref{2.11}) and the equations (\ref{2.47}) imply that the
value
\begin{equation}
\label{2.49}
E = c^{2}(1 - \frac{1}{c^{2}} |\frac{d{\bf x}_{1}}{dt}|^{2})^{- 1/2} -
m_{2}G|{\bf x}_{1}|^{- 1} 
\end{equation}
is constant:
$$
\frac{dE}{dt} = 0.
$$

In view of the relation (\ref{2.48}) the vector ${\bf x}_{1}$ is orthogonal
to the constant vector ${\bf M}$. By making use of the rotation we choose
the coordinate system such that $M_{1} = M_{2} = 0$. Therefore 
$x_{1}^{3} = 0$. Let us choose the polar coordinates
$$
x_{1}^{1}(t) = r(t)\cos \phi (t),
$$
$$
x_{1}^{2}(t) = r(t)\sin \phi (t),
$$
$$
x_{1}^{3}(t) = 0.
$$
The relations (\ref{2.48}), (\ref{2.49}) imply
\begin{equation}
\label{2.50}
M_{3} = r^{2}\frac{d\phi }{dt}
\left( 1 - \frac{1}{c^{2}} \left( \left( \frac{dr}{dt} \right)^{2} +
r^{2}\left( \frac{d\phi }{dt} \right)^{2} \right) \right)^{- 1/2},
\end{equation}
\begin{equation}
\label{2.51}
E = c^{2}\left( 1 - \frac{1}{c^{2}} \left( \left( \frac{dr}{dt} \right)^{2} +
r^{2}\left( \frac{d\phi }{dt} \right)^{2}\right) \right)^{- 1/2} -
m_{2}Gr^{- 1},
\end{equation}
\begin{equation}
\label{2.52}
\frac{d\phi }{dt} = r^{- 2}c^{2}M_{3}(E + m_{2}Gr^{- 1})^{- 1},
\end{equation}
\begin{equation}
\label{2.53}
\left( \frac{dr}{dt} \right)^{2} = c^{2} - c^{6}(E + m_{2}Gr^{- 1})^{- 2}
(1 + r^{- 2}c^{- 2}M_{3}^{2}).
\end{equation}
It follows from the equation (\ref{2.53}) that
\begin{eqnarray}
\label{2.54}
\epsilon \frac{dr}{dt} = c(E + m_{2}Gr^{- 1})^{- 1} 
( (m_{2}^{2}G^{2} - c^{2}M_{3}^{2})
\left( r^{- 1} + m_{2}GE(m_{2}^{2}G^{2} - c^{2}M_{3}^{2})^{- 1}\right)^{2} -
\nonumber \\ 
c^{2}M_{3}^{2}E^{2}(m_{2}^{2}G^{2} - c^{2}M_{3}^{2})^{- 1} - c^{4})^{1/2}
\end{eqnarray}
where $\epsilon = \pm 1$. The equations (\ref{2.52}), (\ref{2.54}) imply
\begin{eqnarray}
\label{2.55}
- \epsilon \left( \frac{d\phi }{dt} \right)^{- 1} \frac{d}{dt} r^{- 1} =
(cM_{3})^{- 1}( (m_{2}^{2}G^{2} - c^{2}M_{3}^{2})
\left( r^{- 1} + m_{2}GE(m_{2}^{2}G^{2} - c^{2}M_{3}^{2})^{- 1}\right)^{2} -
\nonumber \\ 
c^{2}M_{3}^{2}E^{2}(m_{2}^{2}G^{2} - c^{2}M_{3}^{2})^{- 1} - c^{4})^{1/2}
\end{eqnarray}
If
\begin{equation}
\label{2.56}
c^{2}M_{3}^{2} - m_{2}^{2}G^{2} > 0,
\end{equation}
\begin{equation}
\label{2.57}
c^{2}M_{3}^{2}(E^{2} - c^{4}) + m_{2}^{2}G^{2}c^{4} > 0,
\end{equation}
then the solution of the equation (\ref{2.55}) is the following
\begin{equation}
\label{2.58}
\frac{p}{r} = 1 - \epsilon e \sin 
\left( (c^{2}M_{3}^{2} - m_{2}^{2}G^{2})^{1/2} (cM_{3})^{- 1}
(\phi - \phi_{0} )\right)
\end{equation}
where $\phi_{0} $ is a constant and
\begin{equation}
\label{2.59}
p = (c^{2}M_{3}^{2} - m_{2}^{2}G^{2})(m_{2}GE)^{- 1},
\end{equation}
\begin{equation}
\label{2.60}
e = (c^{2}M_{3}^{2}(E^{2} - c^{4}) + m_{2}^{2}G^{2}c^{4})^{1/2}
(m_{2}GE)^{- 1}.
\end{equation}
The equation (\ref{2.58}) is the equation of the conic section with the
focus at the origin of coordinates. For $e^{2} < 1$ this conic section is
an ellipse and the light does not leave the gravitating body.

If the inequality (\ref{2.56}) holds and 
\begin{equation}
\label{2.61}
c^{2} M_{3}^{2} (E^{2} - c^{4}) + m_{2}^{2} G^{2} c^{4} < 0,
\end{equation}
the right hand side of the equation (\ref{2.53}) is negative and the
equation (\ref{2.53}) has no solutions.

If
\begin{equation}
\label{2.62}
m_{2}^{2} G^{2} - c^{2} M_{3}^{2} > 0,
\end{equation}
then the inequality (\ref{2.57}) holds and the equation (\ref{2.55})
has the following solution
\begin{equation}
\label{2.63}
\frac{p}{r} = 1 + \epsilon e \cosh 
\left( (m_{2}^{2} G^{2} - c^{2} M_{3}^{2})^{1/2} (cM_{3})^{- 1}
(\phi - \phi_{0} )\right)
\end{equation}
where $\phi_{0} $ is a constant.

If
\begin{equation}
\label{2.64}
c^{2} M_{3}^{2} - m_{2}^{2} G^{2} = 0,
\end{equation}
the equation (\ref{2.55}) has the following form
\begin{equation}
\label{2.65}
- \epsilon \left( \frac{d\phi }{dt} \right)^{- 1} \frac{d}{dt} r^{- 1} =
(cM_{3})^{- 1}(2m_{2}GEr^{- 1} + E^{2} - c^{4})^{1/2}
\end{equation}
The solution of the equation (\ref{2.65}) is the following
\begin{equation}
\label{2.66}
\phi = - \epsilon M_{3}(|M_{3}|E)^{- 1}
(2m_{2}GEr^{- 1} + E^{2} - c^{4})^{1/2} + \phi_{0}
\end{equation}
where $\phi_{0} $ is a constant.

Let us consider the equation (\ref{2.54}). Let $r$ and $t$ be the function
of the parameter $\xi $. If the inequality (\ref{2.57}) holds and
\begin{equation}
\label{2.67}
E^{2} - c^{4} < 0,
\end{equation}
then the equation (\ref{2.54}) has the following solution
\begin{eqnarray}
\label{2.68}
r(\xi ) = m_{2}GE(c^{4} - E^{2})^{- 1}
(1 + e\sin \left( (c^{4} - E^{2})^{1/2}c^{- 1}(\xi - \xi_{0} )\right) ),
\nonumber \\
t(\xi ) = \epsilon m_{2}GE^{2}c^{- 1}(c^{4} - E^{2})^{- 3/2}
(c^{3}E^{- 2}(c^{4} - E^{2})^{1/2}(\xi - \xi_{1} ) - \nonumber \\
e\cos \left( (c^{4} - E^{2})^{1/2}c^{- 1}(\xi - \xi_{0} )\right) )
\end{eqnarray}
where $\xi_{0} $, $\xi_{1} $ are the constants.

If the inequalities (\ref{2.61}) and (\ref{2.67}) hold, the right hand
side of the equation (\ref{2.53}) is negative and the equation (\ref{2.53})
has no solutions.

If
\begin{equation}
\label{2.69}
E^{2} - c^{4} > 0,
\end{equation}
then the equation (\ref{2.54}) has the following solution
\begin{eqnarray}
\label{2.70}
r(\xi ) = m_{2}GE(E^{2} - c^{4})^{- 1}
(e\cosh \left( (E^{2} - c^{4})^{1/2}c^{- 1}(\xi - \xi_{0} )\right) - 1),
\nonumber \\
t(\xi ) = - \epsilon m_{2}GE^{2}c^{- 1}(E^{2} - c^{4})^{- 3/2}
(c^{3}E^{- 2}(E^{2} - c^{4})^{1/2}(\xi - \xi_{1} ) - \nonumber \\
e\sinh \left( (E^{2} - c^{4})^{1/2}c^{- 1}(\xi - \xi_{0} )\right) )
\end{eqnarray}
where $\xi_{0} $, $\xi_{1} $ are the constants.

If
\begin{equation}
\label{2.71}
E^{2} - c^{4} = 0,
\end{equation}
then the equation (\ref{2.54}) has the following form
\begin{equation}
\label{2.72}
\epsilon \frac{dr}{dt} = c(c^{2}r + m_{2}G)^{- 1}
(2m_{2}Gc^{2}r + m_{2}^{2}G^{2} - c^{2}M_{3}^{2})^{1/2}.
\end{equation}
The equation (\ref{2.72}) has the following solution
\begin{equation}
\label{2.73}
t = \epsilon (2m_{2}Gc^{2}r + m_{2}^{2}G^{2} - c^{2}M_{3}^{2})^{1/2}
(r(3m_{2}Gc)^{- 1} + 2/3c^{- 3} + 1/3M_{3}^{2}(m_{2}^{2}G^{2}c)^{- 1}) 
+ t_{0}
\end{equation}
where $t_{0}$ is a constant.

Let us study the limit of the solutions of the equations (\ref{2.54}),
(\ref{2.55}) when $c \rightarrow \infty $. It follows from the
expressions (\ref{2.48}), (\ref{2.49}) that for $c \rightarrow \infty $
\begin{equation}
\label{2.74}
M_{3} \rightarrow 
x_{1}^{1}\frac{dx_{1}^{2}}{dt} - x_{1}^{2}\frac{dx_{1}^{1}}{dt} =
\bar{M}_{3},
\end{equation}
\begin{equation}
\label{2.75}
c^{- 2}(E^{2} - c^{4}) \rightarrow
|\frac{d{\bf x}_{1}}{dt} |^{2} - 2m_{2}G|{\bf x}_{1}|^{- 1} = 2\bar{E}.
\end{equation}
For $c \rightarrow \infty $ only the inequalities (\ref{2.56}), (\ref{2.57})
are valid. For $c \rightarrow \infty $ the limit of the solution (\ref{2.58})
is the following function
\begin{equation}
\label{2.76}
\frac{\bar{M}_{3}^{2} }{m_{2}G} \frac{1}{r} = 1 -
\epsilon (2\bar{E} \bar{M}_{3}^{2} + m_{2}^{2}G^{2})^{1/2} (m_{2}G)^{- 1}
\sin (\phi - \phi_{0} ).
\end{equation}
Due to (\cite{6}, Chapter III, Section 15) the function (\ref{2.76}) is
the solution of the equations (\ref{2.1}) for $m_{1} = 0$, $x_{2}^{i} = 0$,
$i = 1,2,3$.

For $c \rightarrow \infty $ the functions (\ref{2.68}), (\ref{2.70})
tend to following functions
\begin{eqnarray}
\label{2.77}
r(\xi ) = \frac{m_{2}G}{2|\bar{E} |} (1 + 
(1 + 2\bar{E} \frac{\bar{M}_{3}^{2} }{m_{2}^{2}G^{2}} )^{1/2} 
\sin ((2|\bar{E} |)^{1/2}(\xi - \xi_{0} ))), \nonumber \\
t(\xi ) = \frac{m_{2}G}{(2|\bar{E} |)^{3/2}} ((2|\bar{E} |)^{1/2}
(\xi - \xi_{1} ) - 
(1 + 2\bar{E} \frac{\bar{M}_{3}^{2} }{m_{2}^{2}G^{2}} )^{1/2}
\cos ((2|\bar{E} |)^{1/2}(\xi - \xi_{0} ))) 
\end{eqnarray}
and
\begin{eqnarray}
\label{2.78}
r(\xi ) = \frac{m_{2}G}{2\bar{E} } 
((1 + 2\bar{E} \frac{\bar{M}_{3}^{2} }{m_{2}^{2}G^{2}} )^{1/2} 
\cosh ((2\bar{E} )^{1/2}(\xi - \xi_{0} )) - 1), \nonumber \\
t(\xi ) = - \epsilon \frac{m_{2}G}{(2\bar{E} )^{3/2}} ((2\bar{E} )^{1/2}
(\xi - \xi_{1} ) - 
(1 + 2\bar{E} \frac{\bar{M}_{3}^{2} }{m_{2}^{2}G^{2}} )^{1/2}
\sinh ((2\bar{E} )^{1/2}(\xi - \xi_{0} ))). 
\end{eqnarray}
The formulae (\ref{2.77}), (\ref{2.78}) coincide with the formulae from
(\cite{6}, Chapter III, Section 15) for the equations (\ref{2.1}) when
$m_{1} = 0$, $x_{2}^{i} = 0$, $i = 1,2,3$.

For $c \rightarrow \infty $ the function (\ref{2.73}) tends to the 
following function
\begin{equation}
\label{2.79}
t = \epsilon (2m_{2}Gr - \bar{M}_{3}^{2} )^{1/2}
(r(3m_{2}G)^{- 1} + 1/3\bar{M}_{3}^{2} m_{2}^{- 2}G^{- 2}) + t_{0}.
\end{equation}
This case is not considered in (\cite{6}, Chapter III, Section 15).

Let us study the equations (\ref{2.4}), (\ref{2.41}), (\ref{2.42}) for
the case when the first body is Mercury and the second body is the Sun.
Let us estimate the value of the electromagnetic field strength
$F_{\mu \nu }(x_{2};x_{1})$. We consider that Mercury moves along the
circle of the radius $a$ with the angular frequency $\omega $. Then the
expression (\ref{2.41}) implies
\begin{eqnarray}
\label{2.80}
|F_{i0}(x_{2};x_{1})| \leq m_{1}G(a - a^{2}\omega c^{- 1})^{- 2}a^{2}
(\omega c^{- 1})^{2} + \nonumber \\
m_{1}G(a - a^{2}\omega c^{- 1})^{- 3}(1 + 2a^{2}(\omega c^{- 1})^{2})
(a + a^{2}\omega c^{- 1}),
\end{eqnarray}
\begin{eqnarray}
\label{2.81}
|F_{ij}(x_{2};x_{1})| \leq m_{1}G(a - a^{2}\omega c^{- 1})^{- 2}2a^{2}
(\omega c^{- 1})^{2} + \nonumber \\
m_{1}G(a - a^{2}\omega c^{- 1})^{- 3}(1 + 2a^{2}(\omega c^{- 1})^{2})
2a^{2}\omega c^{- 1}
\end{eqnarray}
where $i,j = 1,2,3$. Due to (\cite{2}, Chapter 25, Section 25.1, 
Appendix 25.1) $m_{1} = 3,28 \cdot 10^{23}\, kg$, $\omega c^{- 1} =
275,8 \cdot 10^{- 17}\, m^{- 1}$, $a = 0,5791 \cdot 10^{11}\, m$. Then
$|F_{i0}(x_{2};x_{1})| \leq 6,52 \cdot 10^{- 9}\, m \cdot s^{- 2}$,
$|F_{i,j}(x_{2},x_{1})| \leq 2,08 \cdot 10^{- 12}\, m \cdot s^{- 2}$.
Therefore the case $m_{1} = 0$ is a good approximation for our problem.

Let us consider the equation (\ref{2.58}). If $e^{2} \geq 1$, there is
the angle $\phi $ for which $r = \infty $. Thus the orbit is not closed.
If $e^{2} < 1$ and $E < 0$, then $p < 0$ and the equation (\ref{2.58})
has no solutions. Let $e^{2} < 1$ and $E > 0$. The substitution
$\epsilon \rightarrow - \epsilon $ corresponds with the substitution
$\phi_{0} \rightarrow \phi_{0} + 
\pi cM_{3}(c^{2}M_{3}^{2} - m_{2}^{2}G^{2})^{- 1/2}$.
We choose $\epsilon = 1$, 
$\phi_{0} = \pi /2cM_{3}(c^{2}M_{3}^{2} - m_{2}^{2}G^{2})^{- 1/2}$.
The curve (\ref{2.58}) is an ellipse with the focus at the origin of
coordinates. Its big and small semiaxes are
\begin{equation}
\label{2.82}
a = p(1 - e^{2})^{- 1} = m_{2}GE(c^{4} - E^{2})^{- 1},
\end{equation}
\begin{equation}
\label{2.83}
b = a(1 - e^{2})^{1/2} = (c^{2}M_{3}^{2} - m_{2}^{2}G^{2})^{1/2}
(c^{4} - E^{2})^{- 1/2} .
\end{equation}
The inequalities (\ref{2.56}) and $e^{2} < 1$ imply the inequality 
(\ref{2.67}). Thus the equations (\ref{2.68}) are valid. For the
parameter $\xi_{2} = \xi_{0} - \pi /2c(c^{4} - E^{2})^{- 1/2}$
\begin{equation}
\label{2.84}
r(\xi_{2} ) = r_{min} = m_{2}GE(c^{4} - E^{2})^{- 1}(1 - e).
\end{equation}
For the parameter $\xi_{3} = \xi_{0} + \pi /2c(c^{4} - E^{2})^{- 1/2}$
\begin{equation}
\label{2.85}
r(\xi_{3} ) = r_{max} = m_{2}GE(c^{4} - E^{2})^{- 1}(1 + e).
\end{equation}
Hence the period of the motion along the ellipse (\ref{2.58}) is equal to
\begin{equation}
\label{2.86}
T = 2|t(\xi_{3} ) - t(\xi_{2} )| = 2\pi m_{2}Gc^{3}(c^{4} - E^{2})^{- 3/2}.
\end{equation}
We define the mean angular frequency $\omega = 2\pi T^{- 1}$. The
relation (\ref{2.86}) implies
\begin{equation}
\label{2.87}
\omega = (c^{4} - E^{2})^{3/2}(m_{2}Gc^{3})^{- 1}.
\end{equation}
Therefore
\begin{equation}
\label{2.88}
E^{2} = c^{2}(c^{2} - (\omega m_{2}G)^{2/3}).
\end{equation}
The substitution of the expression (\ref{2.88}) into the equality 
(\ref{2.82}) yields
\begin{equation}
\label{2.89}
m_{2}G = \omega^{2} a^{3}
\left( 1/2(1 + \sigma (1 - (2a\omega c^{- 1})^{2})^{1/2})\right)^{- 3/2}
\end{equation}
where $\sigma = \pm 1$. Let $c \rightarrow \infty $. Then 
$m_{2}G = \omega^{2} a^{3}(1/2(1 + \sigma ))^{- 3/2}$. For $\sigma = 1$
this expression is in accordance with the expression obtained in the
Kepler problem (\cite{6}, Chapter III, Sections 13, 15). Hence we choose
$\sigma = 1$ in the relation (\ref{2.89}). The substitution of the
expression (\ref{2.89}) with $\sigma = 1$ into the equality (\ref{2.88})
gives
\begin{equation}
\label{2.90}
E^{2} = c^{4} - 2c^{2}\omega^{2} a^{2}
(1 + (1 - (2a\omega c^{- 1})^{2})^{1/2})^{- 1}.
\end{equation}
By making use of the relations (\ref{2.59}), (\ref{2.60}), (\ref{2.82}),
(\ref{2.83}), (\ref{2.89}), (\ref{2.90}) we get
\begin{equation}
\label{2.91}
(c^{2}M_{3}^{2} - m_{2}^{2}G^{2})^{1/2}(c|M_{3}|)^{- 1} =
( 1 + 4a^{2}(\omega c^{- 1})^{2}(1 - e^{2})^{- 1}
(1 + (1 - (2a\omega c^{- 1})^{2})^{1/2})^{- 2})^{- 1/2}.
\end{equation}
The relativistic orbit (\ref{2.58}) differs from the non -- relativistic
orbit (\ref{2.76}) by the multiplier (\ref{2.91}). Let us calculate the
shift of Mercury perihelion for a hundred years in the seconds of arc.
Due to (\cite{2}, Chapter 25, Section 25.1, Appendix 25.1) the period of
Mercury circulation is 87,9686 days and the period of circulation of the
Earth is 365,257 days. Hence
\begin{equation}
\label{2.92}
\Delta \phi = 2\pi (1 - (c^{2}M_{3}^{2} - m_{2}^{2}G^{2})^{1/2}
(c|M_{3}|)^{- 1}) \cdot 100 \cdot 365,257 \cdot (87,9686)^{- 1} \cdot
360 \cdot 3600
\end{equation}
For Mercury $\omega c^{- 1} = 275,8 \cdot 10^{- 17}\, m^{- 1} $,
$a = 0,5791 \cdot 10^{11} \, m$, $e = 0,21$. Therefore the relations
(\ref{2.91}), (\ref{2.92}) imply that $\Delta \phi = 45",09$. Due to
(\cite{2}, Chapter 40, Section 40.5, Appendix 40.3) the shift for a
hundred years which must be attached to the general relativity and to 
the Sun flattening is $\Delta \phi = 42",56 \pm 0",94$.

\section{Lorentz Force}
\setcounter{equation}{0}

\noindent Let the world line $x^{\mu }(\lambda )$, $\mu = 0,...,3$, be
given. It depends on the parameter $\lambda $. Let on ${\bf R}^{4}$
the pseudometric $g_{\mu \nu } = g_{\nu \mu }$ be given. We suppose
that the matrix $g_{\mu \nu }$ has the inverse matrix $g^{\mu \nu }$.
We define the proper time $\tau $ in the following way 
\begin{equation}
\label{3.1}
\frac{d\lambda }{d\tau } = 
\left( \sum_{\mu ,\nu = 0}^{3} g_{\mu \nu}\frac{dx^{\mu}}{d\lambda }
\frac{dx^{\nu }}{d\lambda } \right)^{- 1/2}.
\end{equation}
Due to (\cite{2}, Chapter 13, Section 13.4) the world line 
$x^{\mu }(\lambda )$ is called geodesic if the functional
\begin{equation}
\label{3.2}
\int_{\lambda_{1} }^{\lambda_{2} }
\left( \sum_{\mu ,\nu = 0}^{3} g_{\mu \nu}\frac{dx^{\mu}}{d\lambda }
\frac{dx^{\nu }}{d\lambda } \right)^{1/2} d\lambda
\end{equation}
is extremal. Then the world line $x^{\mu }(\lambda )$ satisfies the
geodesic equation with the affine parameter $\tau $
\begin{equation}
\label{3.3}
\frac{d\lambda }{d\tau } \frac{d}{d\lambda }
\left( \frac{d\lambda }{d\tau } \frac{dx^{\mu }}{d\lambda } \right) +
\sum_{\alpha ,\beta = 0}^{3} \Gamma_{\alpha \beta }^{\mu }
\frac{d\lambda }{d\tau } \frac{dx^{\alpha }}{d\lambda }
\frac{d\lambda }{d\tau } \frac{dx^{\beta }}{d\lambda } = 0
\end{equation}
where the Christoffel's symbol
\begin{equation}
\label{3.4}
\Gamma_{\alpha \beta }^{\mu } = \frac{1}{2} \sum_{\sigma = 0}^{3}
\left( \frac{\partial g_{\sigma \beta }}{\partial x^{\alpha }} +
\frac{\partial g_{\sigma \alpha }}{\partial x^{\beta }} -
\frac{\partial g_{\alpha \beta }}{\partial x^{\sigma }} \right)
g^{\sigma \mu }.
\end{equation}
The equation (\ref{3.3}) does not depend on a choice of a parameter
$\lambda $.

The gravitation interaction of two bodies is given by the Newton gravity
law (\ref{2.1}), (\ref{2.2}). Following (\cite{2}, Chapter 17, Section 17.4)
we shall represent the equations (\ref{2.1}), (\ref{2.2}) as a particular
case of the equations (\ref{3.3}), (\ref{3.4}). Let $U(x;y)$ be the solution
of the equation (\ref{2.2}). It depends on the variables $x^{0} = ct$,
$x^{i}$, $i = 1,2,3$. Let us consider the pseudometric
\begin{eqnarray}
\label{3.5}
g_{00} = 1 + 2c^{- 2}U(x;y), \nonumber \\
g_{i0} = g_{0i} = 0, \nonumber \\
g_{ij} = - \delta_{ij}
\end{eqnarray}
where $i,j = 1,2,3$. Then the definition (\ref{3.1}) implies
\begin{equation}
\label{3.6}
\frac{dt}{d\tau } = c^{- 1}(1 + 2c^{- 2}U - c^{- 2}\sum_{i = 1}^{3}
\left( \frac{dx^{i}}{dt} \right)^{2} )^{- 1/2}.
\end{equation}
It follows from the relations (\ref{3.4}), (\ref{3.5}) that
\begin{eqnarray}
\label{3.7}
\Gamma_{ij}^{\mu } = 0, \nonumber \\
\Gamma_{\mu j}^{i} = \Gamma_{j\mu }^{i} = 0, \nonumber \\
\Gamma_{00}^{0} = c^{- 2}(1 + 2c^{- 2}U)^{- 1}
\frac{\partial U}{\partial x^{0}}, \nonumber \\
\Gamma_{00}^{i} = c^{- 2}\frac{\partial U}{\partial x^{i}}, \nonumber \\
\Gamma_{0i}^{0} = \Gamma_{i0}^{0} = c^{- 2}(1 + 2c^{- 2}U)^{- 1}
\frac{\partial U}{\partial x^{i}}
\end{eqnarray}
where $i,j = 1,2,3$; $\mu = 0,...,3$.

The substitution of the expressions (\ref{3.6}), (\ref{3.7}) into the
equation (\ref{3.3}) for $\mu = 0$ yields 
\begin{eqnarray}
\label{3.8}
c(1 + 2c^{- 2}U - c^{- 2}\sum_{j = 1}^{3}
\left( \frac{dx^{j}}{dt} \right)^{2} )^{1/2}
\frac{d}{dt} (1 + 2c^{- 2}U - c^{- 2}\sum_{j = 1}^{3}
\left( \frac{dx^{j}}{dt} \right)^{2} )^{- 1/2} + \nonumber \\
(1 + 2c^{- 2}U)^{- 1} \left( \frac{\partial U}{\partial x^{0}} +
2c^{- 1}\sum_{j = 1}^{3} \frac{dx^{j}}{dt} 
\frac{\partial U}{\partial x^{j}}\right) = 0.
\end{eqnarray}
The substitution of the expressions (\ref{3.6}), (\ref{3.7}) into the
equations (\ref{3.3}) for $\mu = 1,2,3$ gives
\begin{equation}
\label{3.9}
(1 + 2c^{- 2}U - c^{- 2}\sum_{j = 1}^{3}
\left( \frac{dx^{j}}{dt} \right)^{2} )^{1/2}
\frac{d}{dt} \left( (1 + 2c^{- 2}U - c^{- 2}\sum_{j = 1}^{3}
\left( \frac{dx^{j}}{dt} \right)^{2} )^{- 1/2}
\frac{dx^{i}}{dt} \right) + \frac{\partial U}{\partial x^{i}} = 0
\end{equation}
where $i = 1,2,3$.

We suppose that the velocity $\frac{dx^{i}}{dt}$ is small with respect to
$c$ and the acceleration $\frac{d^{2}x^{i}}{dt^{2}}$ is bounded with
respect to $c$. We also suppose that the potential $U$ is small with
respect to $c$ and its derivatives are bounded with respect to $c$. Then
for $c \rightarrow \infty $ the equations (\ref{3.8}), (\ref{3.9}) get
the following form
\begin{eqnarray}
\label{3.10}
\frac{\partial U}{\partial x^{0}} = 0, \nonumber \\
\frac{d^{2}x^{i}}{dt^{2}} + \frac{\partial U}{\partial x^{i}} = 0.
\end{eqnarray}
The first equation (\ref{3.10}) shows that the potential $U(x;y)$
is independent of the variable $x^{0} = ct$, i.e. the body with the
world line $y^{\mu }$ is at rest. 

Following (\cite{2}, Chapter 12, Section 12.1) we show that the
equations (\ref{2.1}), (\ref{2.2}) are the particular case not of the
geodesic equation (\ref{3.3}) with the affine parameter but of the geodesic
equation with the parameter $t$
\begin{equation}
\label{3.11}
\frac{d^{2}x_{k}^{\mu }}{dt^{2}} +
\sum_{\alpha ,\beta = 0}^{3} 
\Gamma_{\alpha \beta }^{\mu }(x_{k};\hat{x}_{k} )
\frac{dx_{k}^{\alpha }}{dt} \frac{dx_{k}^{\beta }}{dt} = 0
\end{equation}
where $\mu = 0,...,3$, the world lines $x_{k}^{0} = ct$, $x_{k}^{i}(t)$,
$k = 1,2$, $i = 1,2,3$, and the coefficients
\begin{equation}
\label{3.12}
\Gamma_{00}^{i}(x_{k};\hat{x}_{k} ) = c^{- 2}
\frac{\partial U(x_{k};\hat{x}_{k} )}{\partial x_{k}^{i}}.
\end{equation}
All the other coefficients $\Gamma_{\alpha \beta }^{\mu }(x_{k};\hat{x}_{k} )$
are equal to zero.

Let us prove that it is impossible to represent the coefficients (\ref{3.12})
in the form of the Christoffel's symbols (\ref{3.4}). Let these coefficients
have the form (\ref{3.4}). Then
\begin{equation}
\label{3.13}
\frac{1}{2} \left( \frac{\partial g_{\beta \gamma }}{\partial x^{\alpha }} +
\frac{\partial g_{\alpha \gamma }}{\partial x^{\beta }} -
\frac{\partial g_{\alpha \beta }}{\partial x^{\gamma }} \right) =
\sum_{\sigma = 0}^{3} g_{\gamma \sigma } \Gamma_{\alpha \beta }^{\sigma }.
\end{equation}
The cyclic permutation $\alpha \rightarrow \beta \rightarrow \gamma
\rightarrow \alpha $ in the equation (3.13) yields
\begin{equation}
\label{3.14}
\frac{1}{2} \left( \frac{\partial g_{\gamma \alpha }}{\partial x^{\beta }} +
\frac{\partial g_{\beta \alpha }}{\partial x^{\gamma }} -
\frac{\partial g_{\beta \gamma }}{\partial x^{\alpha }} \right) =
\sum_{\sigma = 0}^{3} g_{\alpha \sigma } \Gamma_{\beta \gamma }^{\sigma }.
\end{equation}
In virtue of the symmetry $g_{\nu \mu } = g_{\mu \nu }$ the sum of the
equations (\ref{3.13}), (\ref{3.14}) is
\begin{equation}
\label{3.15}
\frac{\partial g_{\alpha \gamma }}{\partial x^{\beta }} =
\sum_{\sigma = 0}^{3} (g_{\gamma \sigma } \Gamma_{\alpha \beta }^{\sigma } +
g_{\alpha \sigma } \Gamma_{\beta \gamma }^{\sigma } ).
\end{equation}
The equation (\ref{3.15}) and the definition (\ref{3.12}) imply
\begin{equation}
\label{3.16}
\frac{\partial g_{\alpha \gamma }}{\partial x^{i}} = 0
\end{equation}
where $i = 1,2,3$. Hence the pseudometric $g_{\mu \nu }$ does not depend
on the variables $x^{i}$, $i = 1,2,3$, and it is impossible to represent
the coefficients (\ref{3.12}) in the form (\ref{3.4}).

The Newton gravity law seems to be very important. Therefore we propose to
consider the equations of the type (\ref{3.3}) where the coefficients
$\Gamma_{\alpha \beta }^{\mu }$ do not have a form of the Christoffel's
symbols (\ref{3.4}). Thus we propose to refuse the pseudometric 
$g_{\mu \nu }$. We want our equation not to depend on the
parametrization of the world line. Thus we change the pseudometric
$g_{\mu \nu }$ in the expression (\ref{3.1}) for the fixed Minkowski
pseudometric $\eta_{\mu \nu }$
\begin{equation}
\label{3.17}
\frac{d\lambda }{ds} = 
\left( \sum_{\mu ,\nu = 0}^{3} \eta_{\mu \nu} \frac{dx^{\mu}}{d\lambda }
\frac{dx^{\nu }}{d\lambda } \right)^{- 1/2}.
\end{equation}
By changing the expression (\ref{3.1}) in the equation (\ref{3.3}) for
the expression (\ref{3.17}) we get
\begin{equation}
\label{3.18}
\frac{d\lambda }{ds} \frac{d}{d\lambda }
\left( \frac{d\lambda }{ds} \frac{dx^{\mu }}{d\lambda } \right) +
\sum_{\alpha ,\beta = 0}^{3} \Gamma_{\alpha \beta }^{\mu }
\frac{d\lambda }{ds} \frac{dx^{\alpha }}{d\lambda }
\frac{d\lambda }{ds} \frac{dx^{\beta }}{d\lambda } = 0
\end{equation}
The equation (\ref{3.18}) does not depend on a choice of a parameter
$\lambda $. We choose the time $t$ as the parameter. We want our
equation to be Lagrangian as the equation (\ref{3.3}). Thus we consider
the whole class of the equations of the type (\ref{3.18})
\begin{equation}
\label{3.19}
mc\frac{dt}{ds} \frac{d}{dt}
\left( \frac{dt}{ds} \frac{dx^{\mu }}{dt} \right) +
\frac{e}{c} \sum_{k = 0}^{N} \sum_{\alpha_{1},...,\alpha_{k} = 0}^{3} 
\eta^{\mu \mu } F_{\mu \alpha_{1} \cdots \alpha_{k} }(x)
\frac{dt}{ds} \frac{dx^{\alpha_{1} }}{dt} \cdots
\frac{dt}{ds} \frac{dx^{\alpha_{k} }}{dt} = 0
\end{equation}
and choose the Lagrangian equations. We consider the world line 
$x^{\mu }(t)$: $x^{0}(t) = ct$, $x^{i}(t)$, $i = 1,2,3$. The definitions
(\ref{2.7}), (\ref{2.8}) imply
\begin{equation}
\label{3.20}
\sum_{\alpha = 0}^{3} \eta_{\alpha \alpha }
\left( \frac{dx^{\alpha }}{ds} \right)^{2} = 1,
\end{equation}
\begin{equation}
\label{3.21}
\sum_{\alpha = 0}^{3} \eta_{\alpha \alpha }
\frac{dx^{\alpha }}{ds} \frac{d^{2}x^{\alpha }}{ds^{2}} = 0.
\end{equation}
The identity (\ref{3.21}) coincides with the identity (\ref{2.11}).

The equation (\ref{3.19}) and the identity (\ref{3.21}) imply
\begin{equation}
\label{3.22}
\sum_{k = 0}^{N} \sum_{\mu, \alpha_{1},...,\alpha_{k} = 0}^{3} 
F_{\mu \alpha_{1} \cdots \alpha_{k} }(x)
\frac{dx^{\mu }}{ds}
\frac{dx^{\alpha_{1} }}{ds} \cdots
\frac{dx^{\alpha^{k} }}{ds} = 0
\end{equation}
Conversely, it follows from the equations (\ref{3.21}), (\ref{3.22})
that the equation (\ref{3.19}) for $\mu = 0$ is the cosequence of the
equations (\ref{3.19}) for $\mu = 1,2,3$. Let us denote 
$v^{i} = \frac{dx^{i}}{dt} $, $i = 1,2,3$. The substitution of the 
expression (\ref{2.8}) into the equation (\ref{3.19}) gives
\begin{eqnarray}
\label{3.23}
m\frac{d}{dt} 
\left( (1 - \frac{1}{c^{2}} |{\bf v}|^{2})^{- 1/2} v^{i} \right) -
\nonumber \\
e\sum_{k = 0}^{N} c^{- k}(1 - \frac{1}{c^{2}} |{\bf v}|^{2})^{- \frac{k - 1}{2}}
\sum_{\alpha_{1},...,\alpha_{k} = 0}^{3} 
F_{i \alpha_{1} \cdots \alpha_{k} }(x)
\frac{dx^{\alpha_{1} }}{dt} \cdots \frac{dx^{\alpha_{k} }}{dt} = 0.
\end{eqnarray}
{\bf Lemma 3.1.} {\it Let there exist the Lagrange function} 
$L({\bf x},{\bf v},t)$ {\it such that for any world line} $x^{\mu }(t)$,
$x^{0}(t) = ct$, {\it and for any} $i = 1,2,3$
\begin{eqnarray}
\label{3.24}
\frac{d}{dt} \frac{\partial L}{\partial v^{i}} -
\frac{\partial L}{\partial x^{i}} = 
m\frac{d}{dt} 
\left( (1 - \frac{1}{c^{2}} |{\bf v}|^{2})^{- 1/2} v^{i} \right) -
\nonumber \\
e\sum_{k = 0}^{N} c^{- k}
(1 - \frac{1}{c^{2}} |{\bf v}|^{2})^{- \frac{k - 1}{2}}
\sum_{\alpha_{1},...,\alpha_{k} = 0}^{3} 
F_{i \alpha_{1} \cdots \alpha_{k} }(x)
\frac{dx^{\alpha_{1} }}{dt} \cdots \frac{dx^{\alpha_{k} }}{dt}. 
\end{eqnarray}
{\it Then the Lagrange function has the form}
\begin{equation}
\label{3.25}
L({\bf x},{\bf v},t) = - mc^{2}(1 - \frac{1}{c^{2}} |{\bf v}|^{2})^{1/2} +
\frac{e}{c} \sum_{i = 1}^{3} A_{i}({\bf x},t)v^{i} + eA_{0}({\bf x},t),
\end{equation}
{\it the coefficients in the equation} (\ref{3.19}) {\it are}
\begin{equation}
\label{3.26}
F_{i \alpha_{1} \cdots \alpha_{k} }(x) = 0, \, \, \, k \neq 1
\end{equation}
{\it and}
\begin{eqnarray}
\label{3.27}
F_{ij}(x) = \frac{\partial A_{j}({\bf x},t)}{\partial x^{i}} -
\frac{\partial A_{i}({\bf x},t)}{\partial x^{j}}, \nonumber \\
F_{i0}(x) = \frac{\partial A_{0}({\bf x},t)}{\partial x^{i}} -
\frac{1}{c} \frac{\partial A_{i}({\bf x},t)}{\partial t}
\end{eqnarray}
{\it where} $i,j = 1,2,3$; $\alpha_{1} ,...,\alpha_{k} = 0,...,3$.

\noindent {\it Proof.} We look for the Lagrange function in the following
form
\begin{equation}
\label{3.28}
L({\bf x},{\bf v},t) = - mc^{2}(1 - \frac{1}{c^{2}} |{\bf v}|^{2})^{1/2} +
L_{1}({\bf x},{\bf v},t).
\end{equation}
The substitution of the expression (\ref{3.28}) into the equality 
(\ref{3.24}) yields
\begin{eqnarray}
\label{3.29}
\frac{d}{dt} \frac{\partial L_{1}}{\partial v^{i}} -
\frac{\partial L_{1}}{\partial x^{i}} = 
\nonumber \\
- e\sum_{k = 0}^{N} c^{- k}
(1 - \frac{1}{c^{2}} |{\bf v}|^{2})^{- \frac{k - 1}{2}}
\sum_{\alpha_{1},...,\alpha_{k} = 0}^{3} 
F_{i \alpha_{1} \cdots \alpha_{k} }(x)
\frac{dx^{\alpha_{1} }}{dt} \cdots \frac{dx^{\alpha_{k} }}{dt}. 
\end{eqnarray}
The right hand side of the equality (\ref{3.29}) has not the vector
$\frac{dv^{i}}{dt} = \frac{d^{2}x^{i}}{dt^{2}} $. Since the world line
$x^{i}(t)$, $i = 1,2,3$, is arbitrary, the left hand side of the equality
(\ref{3.29}) can not have the vector $\frac{dv^{i}}{dt} $, i.e. the 
function $L_{1}({\bf x},{\bf v},t)$ is linear for the variables $v^{i}$
\begin{equation}
\label{3.30}
L_{1}({\bf x},{\bf v},t) = 
\frac{e}{c} \sum_{i = 1}^{3} A_{i}({\bf x},t)v^{i} + eA_{0}({\bf x},t).
\end{equation}
By definition
\begin{equation}
\label{3.31}
\frac{dA_{\mu }({\bf x},t)}{dt} = \sum_{j = 1}^{3}
\frac{\partial A_{\mu }({\bf x},t)}{\partial x^{j}} v^{j} +
\frac{\partial A_{\mu }({\bf x},t)}{\partial t}
\end{equation}
where $\mu = 0,...,3$. The substitution of the expression (\ref{3.31})
into the left hand side of the equality (\ref{3.29}) gives
\begin{eqnarray}
\label{3.32}
\frac{1}{c} \sum_{j = 1}^{3} 
\left( \frac{\partial A_{j}({\bf x},t)}{\partial x^{i}} -
\frac{\partial A_{i}({\bf x},t)}{\partial x^{j}} \right) v^{j} +
\frac{\partial A_{0}({\bf x},t)}{\partial x^{i}} -
\frac{1}{c} \frac{\partial A_{i}({\bf x},t)}{\partial t} = \nonumber \\
\sum_{k = 0}^{N} c^{- k}
(1 - \frac{1}{c^{2}} |{\bf v}|^{2})^{- \frac{k - 1}{2}}
\sum_{\alpha_{1},...,\alpha_{k} = 0}^{3} 
F_{i \alpha_{1} \cdots \alpha_{k} }(x)
\frac{dx^{\alpha_{1} }}{dt} \cdots \frac{dx^{\alpha_{k} }}{dt}.
\end{eqnarray}
We note that $\frac{dx^{0}}{dt} = c$, $\frac{dx^{i}}{dt} = v^{i}$,
$i = 1,2,3$. Since the world line $x^{i}(t)$, $i = 1,2,3$, is arbitrary,
the equation (\ref{3.32}) implies the equalities (\ref{3.26}),
(\ref{3.27}). The lemma is proved.

Let us define 
\begin{equation}
\label{3.33}
F_{00} = 0,
\end{equation}
\begin{equation}
\label{3.34}
F_{0i} = - F_{i0}
\end{equation}
where $i = 1,2,3$. Then the following identity is valid
\begin{equation}
\label{3.35}
\sum_{\alpha, \beta = 0}^{3} F_{\alpha \beta } \frac{dx^{\alpha }}{ds}
\frac{dx^{\beta }}{ds} = 0
\end{equation}
by virtue of the antisymmetry $F_{\beta \alpha } = - F_{\alpha \beta}$.
The equation (\ref{3.35}) is a particular case of the equation 
(\ref{3.22}). Thus we obtain the equation
\begin{equation}
\label{3.36}
mc\frac{d^{2}x^{\mu }}{ds^{2}} = - \frac{e}{c} \eta^{\mu \mu}
\sum_{\nu = 0}^{3} F_{\mu \nu }(x)\frac{dx^{\nu }}{ds}
\end{equation}
where $F_{\mu \nu }(x)$ are given by the equalities (\ref{3.27}),
(\ref{3.33}), (\ref{3.34}).

The equation (\ref{3.36}) is well -- known (\cite{2}, Chapter 3, Section 3.1;
\cite{5}, Chapter III, Section 23). The right hand side of the equation 
(\ref{3.36}) divided by $\frac{dt}{ds} $ is called the Lorentz force.

\section{Connection}
\setcounter{equation}{0}

\noindent The affine connection at the point $x \in {\bf R}^{4}$ is the
function establishing a correspondence between the tangent vector
$X \in T{\bf R}_{x}^{4}$, the vector field $Y$ and a new tangent
vector $\nabla_{X} Y \in T{\bf R}_{x}^{4}$ called the covariant derivative
of $Y$ in the direction $X$. This vector is required to be a
bilinear function of $X$ and $Y$. Moreover, if $f(x)$ is a real function
and if $fY$ is the vector field $(fY)_{y} = f(y)Y_{y}$, then 
the operation $\nabla $ is required to satisfy the condition
\begin{equation}
\label{4.1}
\nabla_{X} (fY) = (Xf)Y_{x} + f(x)\nabla_{X} Y.
\end{equation}
The affine connection (or simply connection) on ${\bf R}^{4}$ is the
function establishing a correspondence between the point 
$x \in {\bf R}^{4}$ and the affine connection at the point $x \in {\bf R}^{4}$
such that the following smooth condition is fulfilled: If $X$ and $Y$ are
the smooth vector fields on ${\bf R}^{4}$, then the vector field
$(\nabla_{X} Y)_{x} = \nabla_{X_{x}} Y$ must be smooth.

Let us denote $\partial_{\mu } $ the vector field 
$\frac{\partial }{\partial x^{\mu }} $ on ${\bf R}^{4}$ where $x^{\mu }$
is a coordinate of the point $x \in {\bf R}^{4}$. Any vector field $U$
on ${\bf R}^{4}$ has the form
\begin{equation}
\label{4.2}
U = \sum_{\mu = 0}^{3} u^{\mu }(x)\frac{\partial }{\partial x^{\mu }}.
\end{equation}
In particular
\begin{equation}
\label{4.3}
\nabla_{\partial_{\mu } } \partial_{\nu } = \sum_{\lambda = 0}^{3}
A_{\mu \nu }^{\lambda }(x)\frac{\partial }{\partial x^{\lambda }}.
\end{equation}
The relation (\ref{4.1}) and the bilinearity of $\nabla_{U} V$ imply
\begin{eqnarray}
\label{4.4}
\nabla_{U} V = \sum_{\lambda = 0}^{3}
\left( \sum_{\mu = 0}^{3} u^{\mu }v_{,\, \mu }^{\lambda }\right)
\frac{\partial }{\partial x^{\lambda }}, \nonumber \\
v_{,\, \mu }^{\lambda } = \frac{\partial v^{\lambda }}{\partial x^{\mu }} +
\sum_{\nu = 0}^{3} A_{\mu \nu }^{\lambda}v^{\nu } .
\end{eqnarray}
Let us introduce the new coordinates $y^{\mu }(x)$
\begin{eqnarray}
\label{4.5}
\frac{\partial }{\partial x^{\mu }} = \sum_{\nu = 0}^{3} C_{\mu}^{\nu}
\frac{\partial }{\partial y^{\nu }}, \nonumber \\
C_{\mu}^{\nu} = \frac{\partial y^{\nu}(x)}{\partial x^{\mu }}.
\end{eqnarray}
The relations (\ref{4.3}) -- (\ref{4.5}) imply
\begin{equation}
\label{4.6}
\sum_{\lambda, \kappa = 0}^{3} A_{\mu \nu}^{\lambda}(x)C_{\lambda}^{\kappa}
\frac{\partial }{\partial y^{\kappa }} = 
\nabla_{(C\frac{\partial }{\partial y})_{\mu}}
(C\frac{\partial }{\partial y})_{\nu} =
\sum_{\lambda, \kappa = 0}^{3} C_{\mu}^{\lambda}
\left( \frac{\partial C_{\nu}^{\kappa}}{\partial y^{\lambda }} +
\sum_{\sigma = 0}^{3} A_{\lambda \sigma}^{\kappa}(y)C_{\nu}^{\sigma}\right)
\frac{\partial }{\partial y^{\kappa }},
\end{equation}
\begin{equation}
\label{4.7}
A_{\mu \nu}^{\lambda}(x) = \sum_{\kappa, \sigma, \tau = 0}^{3}
C_{\mu}^{\tau}C_{\nu}^{\sigma}(C^{- 1})_{\kappa}^{\lambda}
A_{\tau \sigma}^{\kappa}(y) +
\sum_{\kappa = 0}^{3} \frac{\partial C_{\nu}^{\kappa}}{\partial x^{\mu }}
(C^{- 1})_{\kappa}^{\lambda}.
\end{equation}
Thus, if $y^{\mu}(x)$ is a linear function of the variables $x^{\nu}$,
then due to the relation (\ref{4.7}) the connection coefficients
$A_{\mu \nu}^{\lambda}(x)$ transform as a tensor.

Let us define the torsion
\begin{equation}
\label{4.8}
T(X,Y) = \nabla_{X} Y - \nabla_{Y} X - [X,Y],
\end{equation}
\begin{equation}
\label{4.9}
[X,Y]f = X(Yf) - Y(Xf).
\end{equation}
A torsion is a bilinear operation with respect to the multiplication 
by a function: If the vector fields $U$,$V$ have the form (\ref{4.2}),
then
\begin{equation}
\label{4.10}
T(U,V) = \sum_{\lambda, \mu, \nu = 0}^{3} T_{\mu \nu}^{\lambda} u^{\mu}
v^{\nu}\frac{\partial }{\partial x^{\lambda }},
\end{equation}
\begin{equation}
\label{4.11}
T_{\mu \nu}^{\lambda}(x) = A_{\mu \nu}^{\lambda}(x) - 
A_{\nu \mu}^{\lambda}(x).
\end{equation}
A connection with zero torsion is called symmetric. Due to the relation
(\ref{4.11}) for a symmetric connection
\begin{equation}
\label{4.12}
A_{\mu \nu}^{\lambda}(x) = A_{\nu \mu}^{\lambda}(x).
\end{equation}

Let us define the curvature
\begin{equation}
\label{4.13}
R(X,Y)Z = [\nabla_{X}, \nabla_{Y} ]Z - \nabla_{[X,Y]} Z.
\end{equation}
A curvature is a trilinear operation with respect to the multiplication
by a function: If the vector fields $U$,$V$,$W$ have the form (\ref{4.2}),
then
\begin{equation}
\label{4.14}
R(U,V)W = \sum_{\kappa, \lambda, \mu, \nu = 0}^{3} u^{\kappa}v^{\lambda}
w^{\mu}R_{\mu \kappa \lambda}^{\nu}\frac{\partial }{\partial x^{\nu}},
\end{equation}
\begin{equation}
\label{4.15}
R_{\mu \kappa \lambda}^{\nu}(x) = 
\frac{\partial A_{\lambda \mu}^{\nu}(x)}{\partial x^{\kappa}} -
\frac{\partial A_{\kappa \mu}^{\nu}(x)}{\partial x^{\lambda}} +
\sum_{\sigma = 0}^{3} (A_{\lambda \mu}^{\sigma}(x)A_{\kappa \sigma}^{\nu}(x)
- A_{\kappa \mu}^{\sigma}(x)A_{\lambda \sigma}^{\nu}(x)).
\end{equation}
The equality (\ref{4.14}) is proved in (\cite{7}, Lemma 9.1).

Let us introduce the new coordinates $y^{\mu}(x)$ and define the matrix
(\ref{4.5}). The relation (\ref{4.14}) implies
\begin{equation}
\label{4.16}
R_{\mu \kappa \lambda}^{\nu}(x) = \sum_{\alpha, \beta, \gamma, \delta = 0}^{3}
C_{\mu}^{\alpha}C_{\kappa}^{\beta}C_{\lambda}^{\gamma}(C^{- 1})_{\delta}^{\nu}
R_{\alpha \beta \gamma}^{\delta}(y). 
\end{equation}
Thus the curvature (\ref{4.15}) is a tensor for any invertible smooth
functions $y^{\mu}(x)$. 

Let us introduce $4\times 4$ matrices
\begin{equation}
\label{4.17}
(\hat{A}_{\kappa} )_{\mu}^{\nu} = A_{\mu \kappa}^{\nu}(x),
\end{equation}
\begin{equation}
\label{4.18}
(\hat{R}_{\kappa \lambda} )_{\mu}^{\nu} = R_{\mu \kappa \lambda}^{\nu}(x).
\end{equation}
For a symmetric connection the equality (\ref{4.15}) may be rewritten as
\begin{equation}
\label{4.19}
\hat{R}_{\kappa \lambda} = 
\frac{\partial }{\partial x^{\kappa}} \hat{A}_{\lambda} -
\frac{\partial }{\partial x^{\lambda}} \hat{A}_{\kappa} -
[\hat{A}_{\kappa}, \hat{A}_{\lambda} ]
\end{equation}
where the commutator of two matrices
\begin{equation} 
\label{4.20}
([\hat{A}, \hat{B} ])_{\mu}^{\nu} = 
(\hat{A} \hat{B} - \hat{B} \hat{A} )_{\mu}^{\nu} =
\sum_{\sigma = 0}^{3} (\hat{A}_{\mu}^{\sigma} \hat{B}_{\sigma}^{\nu} -
\hat{B}_{\mu}^{\sigma} \hat{A}_{\sigma}^{\nu} ).
\end{equation}
Thus the curvature (\ref{4.15}) looks like a field strength for the
Yang -- Mills field.

Let us define
\begin{equation}
\label{4.21}
(\nabla_{Z} R)(X,Y) = [\nabla_{Z}, R(X,Y)] - R(\nabla_{Z} X,Y) -
R(X,\nabla_{Z} Y).
\end{equation}
For a symmetric connection the following identities are valid:

\noindent the first Bianchi's identity
\begin{equation}
\label{4.22}
R(X,Y)Z + R(Y,Z)X + R(Z,X)Y = 0,
\end{equation}
the second Bianchi's identity
\begin{equation}
\label{4.23}
(\nabla_{Z} R)(X,Y) + (\nabla_{Y} R)(Z,X) + (\nabla_{X} R)(Y,Z) = 0.
\end{equation}
The identities (\ref{4.22}), (\ref{4.23}) are proved in (\cite{8},
Chapter 3, Section 6, Theorems 1,2).

Let on ${\bf R}^{4}$ the pseudometric be given. We denote by
$\langle X,Y\rangle $ the bilinear form of the two vectors 
$X,Y \in T{\bf R}_{x}^{4}$. We suppose that $\langle Y,X\rangle =
\langle X,Y\rangle $ and the matrix $g_{\mu \nu} = 
\langle \frac{\partial }{\partial x^{\mu}}, 
\frac{\partial }{\partial x^{\nu}} \rangle $ is invertible. Its inverse
matrix is denoted by $g^{\mu \nu}$. A connection is called metric if
\begin{equation}
\label{4.24}
X\langle Y,Z\rangle = \langle \nabla_{X} Y,Z\rangle + 
\langle Y,\nabla_{X} Z\rangle
\end{equation}
for any vector fields $X,Y,Z$.

${\bf R}^{4}$ with pseudometric $g_{\mu \nu}$ has the unique metric
connection with zero torsion
\begin{equation}
\label{4.25}
A_{\mu \nu}^{\lambda} = \frac{1}{2} \sum_{\kappa = 0}^{3}
\left( \frac{\partial g_{\nu \kappa}}{\partial x^{\mu}} +
\frac{\partial g_{\mu \kappa}}{\partial x^{\nu}} -
\frac{\partial g_{\mu \nu}}{\partial x^{\kappa}} \right) g^{\kappa \lambda}.
\end{equation}
This proposition is called the basic lemma of Riemann geometry (\cite{7},
Lemma 8.6). The expression (\ref{4.25}) coincides with the Christoffel's
symbol (\ref{3.4}). The metric connection with zero torsion is called the
Riemannian connection (sometimes it is given the name of Levi -- Civita 
connection). It was shown in Section 3 that an arbitrary symmetric 
connection is not a Riemannian connection.

\noindent {\bf Lemma 4.1.} {\it Let on} ${\bf R}^{4}$ {\it the symmetric
connection with the coefficients} $A_{\mu \nu}^{\lambda}$ {\it be given.
By using the curvature coefficients} (\ref{4.15}) {\it we define two tensors:}

\noindent {\it the Ricci tensor}
\begin{equation}
\label{4.26}
R_{\mu \nu} = \sum_{\kappa = 0}^{3} R_{\nu \kappa \mu}^{\kappa} = -
\sum_{\kappa = 0}^{3} R_{\nu \mu \kappa}^{\kappa}
\end{equation}
{\it and the tensor}
\begin{equation}
\label{4.27}
F_{\mu \nu} = \sum_{\kappa = 0}^{3} R_{\kappa \mu \nu}^{\kappa}.
\end{equation}
{\it Then}
\begin{equation}
\label{4.28}
R_{\mu \nu} = \sum_{\kappa = 0}^{3} 
\frac{\partial A_{\mu \nu}^{\kappa}}{\partial x^{\kappa}} -
\frac{\partial }{\partial x^{\mu}} 
\left( \sum_{\kappa = 0}^{3} A_{\kappa \nu}^{\kappa} \right) +
\sum_{\kappa, \sigma = 0}^{3} 
(A_{\mu \nu}^{\sigma}A_{\kappa \sigma}^{\kappa}
- A_{\kappa \nu}^{\sigma}A_{\sigma \mu}^{\kappa}),
\end{equation}
\begin{equation}
\label{4.29}
F_{\mu \nu} = \frac{\partial }{\partial x^{\mu}} 
\left( \sum_{\kappa = 0}^{3} A_{\kappa \nu}^{\kappa} \right) -
\frac{\partial }{\partial x^{\nu}} 
\left( \sum_{\kappa = 0}^{3} A_{\kappa \mu}^{\kappa} \right), 
\end{equation}
\begin{equation}
\label{4.30}
R_{\nu \mu} - R_{\mu \nu} = F_{\mu \nu}.
\end{equation}

{\it If the symmetric connection with the coefficients} 
$A_{\mu \nu}^{\lambda}$ {\it is Riemannian, then}
\begin{equation}
\label{4.31}
F_{\mu \nu} = 0,
\end{equation}
\begin{equation}
\label{4.32}
R_{\nu \mu} = R_{\mu \nu}.
\end{equation}
{\it Proof.} The relation (\ref{4.15}) implies the relations (\ref{4.28}) --
(\ref{4.30}).

If the connection is Riemannian, then the coefficients 
$A_{\mu \nu}^{\lambda}$   have the form (\ref{4.25}) and
\begin{equation}
\label{4.33}
\sum_{\kappa = 0}^{3} A_{\kappa \mu}^{\kappa} = \frac{1}{2}
\sum_{\kappa, \lambda = 0}^{3} g^{\kappa \lambda}
\frac{\partial g_{\kappa \lambda}}{\partial x^{\mu}}.
\end{equation}
The  definition of the inverse matrix $g^{\kappa \lambda}$ implies
\begin{equation}
\label{4.34}
\frac{\partial g^{\kappa \lambda}}{\partial x^{\mu}} = -
\sum_{\sigma, \tau = 0}^{3} g^{\kappa \sigma}g^{\lambda \tau}
\frac{\partial g_{\sigma \tau}}{\partial x^{\mu}}.
\end{equation}
It follows from the relations (\ref{4.33}), (\ref{4.34}) that
\begin{equation}
\label{4.35}
\frac{\partial }{\partial x^{\nu}} 
\left( \sum_{\kappa = 0}^{3} A_{\kappa \mu}^{\kappa} \right) =
\frac{1}{2} \sum_{\kappa, \lambda = 0}^{3} g^{\kappa \lambda}
\frac{\partial^{2} g_{\kappa \lambda}}{\partial x^{\mu}\partial x^{\nu}} -
\frac{1}{2} \sum_{\kappa, \lambda, \sigma, \tau = 0}^{3} 
g^{\kappa \sigma}g^{\lambda \tau}
\frac{\partial g_{\kappa \lambda}}{\partial x^{\mu}}
\frac{\partial g_{\sigma \tau}}{\partial x^{\nu}}.
\end{equation}
The substitution of the relation (\ref{4.35}) into the relation 
(\ref{4.29}) yields the relation (\ref{4.31}). The relations (\ref{4.30}),
(\ref{4.31}) imply the relation (\ref{4.32}). The lemma is proved.

The tensor (\ref{4.29}) has the form of the eletromagnetic field strength
tensor. Let the connection coefficients satisfy the equation
\begin{eqnarray}
\label{4.36}
\sum_{\alpha = 0}^{3} \eta^{\alpha \alpha}
\left( \frac{\partial }{\partial x^{\alpha}} \right)^{2}
A_{\mu \nu}^{\lambda}(x;y) = 4\pi m_{y}\eta_{\mu \mu} \eta_{\nu \nu} 
\frac{dy^{\mu}(\frac{x^{0}}{c} )}{dx^{0}}
\frac{dy^{\nu}(\frac{x^{0}}{c} )}{dx^{0}}
\frac{dy^{\lambda}(\frac{x^{0}}{c} )}{dx^{0}} \times \nonumber \\
\left( \sum_{\alpha = 0}^{3} \eta_{\alpha \alpha} 
\left( \frac{dy^{\alpha}(\frac{x^{0}}{c} )}{dx^{0}} \right)^{2} \right)^{- 1}
\delta ({\bf x} - {\bf y}(\frac{x^{0}}{c} )).
\end{eqnarray}
Then the equations (\ref{4.29}), (\ref{4.36}) imply the equations 
(\ref{2.5}), (\ref{2.6}).

\end{document}